# In-plane anisotropic faceting of ultralarge and thin single-crystalline colloidal SnS nanosheets

Fu Li,[1] Mohammad Mehdi Ramin Moayed,[1]

Eugen Klein,[1] Rostyslav Lesyuk,[1,2] Christian Klinke[1,3,*]

*[1] Institute of Physical Chemistry, University of Hamburg,*
*Martin-Luther-King-Platz 6, 20146 Hamburg, Germany*

*[2] Pidstryhach Institute for applied problems of mechanics and mathematics of NAS of Ukraine, Naukowa str. 3b,*
*79060 Lviv, Ukraine*

*[3] Department of Chemistry, Swansea University - Singleton Park,*
*Swansea SA2 8PP, United Kingdom*

**\*** Corresponding author: christian.klinke@swansea.ac.uk

**Abstract:** The colloidal synthesis of large thin two-dimensional (2D) nanosheets is fascinating but challenging, since the growth along the lateral and vertical dimensions need to be controlled independently. In-plane anisotropy in 2D nanosheets is attracting more attention as well. We present a new synthesis for large colloidal single-crystalline SnS nanosheets with the thicknesses down to 7 nm and lateral sizes up to 8 µm. The synthesis uses trioctylphosphine-S (TOP-S) as sulfur source and oleic acid (with or without TOP) as ligands. Upon adjusting the capping ligand amount, the growth direction can be switched between anisotropic directions (armchair and zigzag) and isotropic directions ("ladder" directions), leading to an edge-morphology anisotropy. This is the first report on solution-phase synthesis of large thin SnS NSs with tunable edge faceting. Furthermore, electronic transport measurements show strong dependency on the crystallographic directions confirming structural anisotropy.

## Table of content

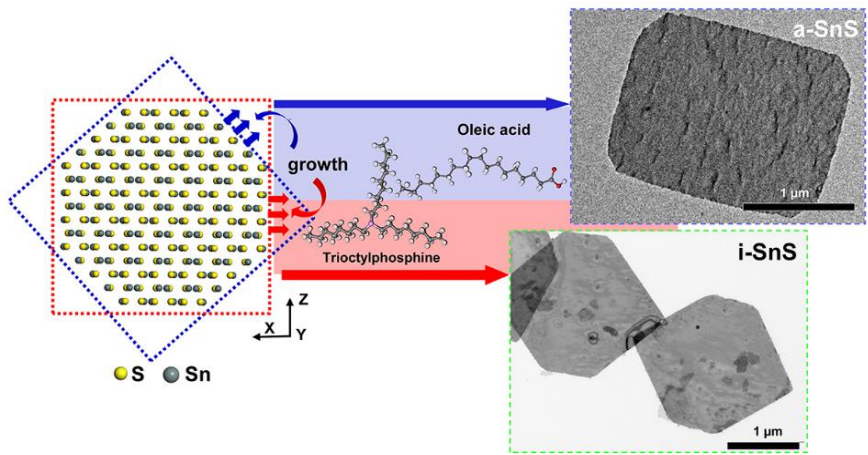

**Keywords:** 2D nanocrystals, SnS nanosheets, anisotropic faceting, electronic anisotropy

Intriguing in-plane anisotropy is now drawing more attention on a couple of 2D nanocrystals (e.g. phosphorene, $ReS_2$), compared to conventional in-plane isotropic materials with high in-plane symmetry (e.g. graphene, $MoS_2$).[1-3] The investigation of in-plane anisotropy can promote the optimization of device design, as well as the enhancement of practical device performance, according to differed responses to external conditions like polarized light or an electric field, opening new possibilities for electronic and optoelectronic devices.[4-5] In particular, 2D tin (II) sulfide (SnS), a phosphorene analogue, is considered as promising anisotropic 2D nanomaterial.[6-7] SnS has a typically layered orthorhombic (OR) crystal structure (*space group: Pbnm*) and can be described as atomic double-layered distorted rock salt structure with in-plane covalent bonds and van-der-Waals bonds vertically along the $b$ direction ($a$=4.33 Å, $b$=11.19 Å, $c$=3.98 Å).[7-10] Therefore, the 2D morphology is preferably obtained for this material with a chemically inert surface without dangling bonds, showing SnS could be a promising alternative for 2D nanomaterial applications.[10]

Colloidal synthesis has been demonstrated to be a facile and highly tunable method for SnS nanomaterials.[11-14] However, a facile method for large-sized SnS with small thickness for electronic applications is still missing. So far, only Ching-Ping Wong's group has reported to produce large-size, high-quality SnS thin single nanocrystals by a high-pressure solvothermal method.[15] We report a new synthesis of single crystal OR SnS NSs by the hot-injection colloidal method with thicknesses down to 7 nm and lateral sizes up to 8 μm, which is the first time for the colloidal synthesis of SnS NSs with this limit of size and thickness. Furthermore, this is also the first instance for a solution-phase synthesis of large thin SnS NSs with tunable edge faceting (edges parallel to anisotropic armchair/zigzag directions, or parallel to the isotropic "ladder" directions). Our results of the application for direction-dependent electrical transport show that the conductivity along the zigzag direction is higher than the armchair direction, and the [101] and the [10-1] direction do not show significant differences, confirming the (an-)isotropy of NSs' crystal structure.

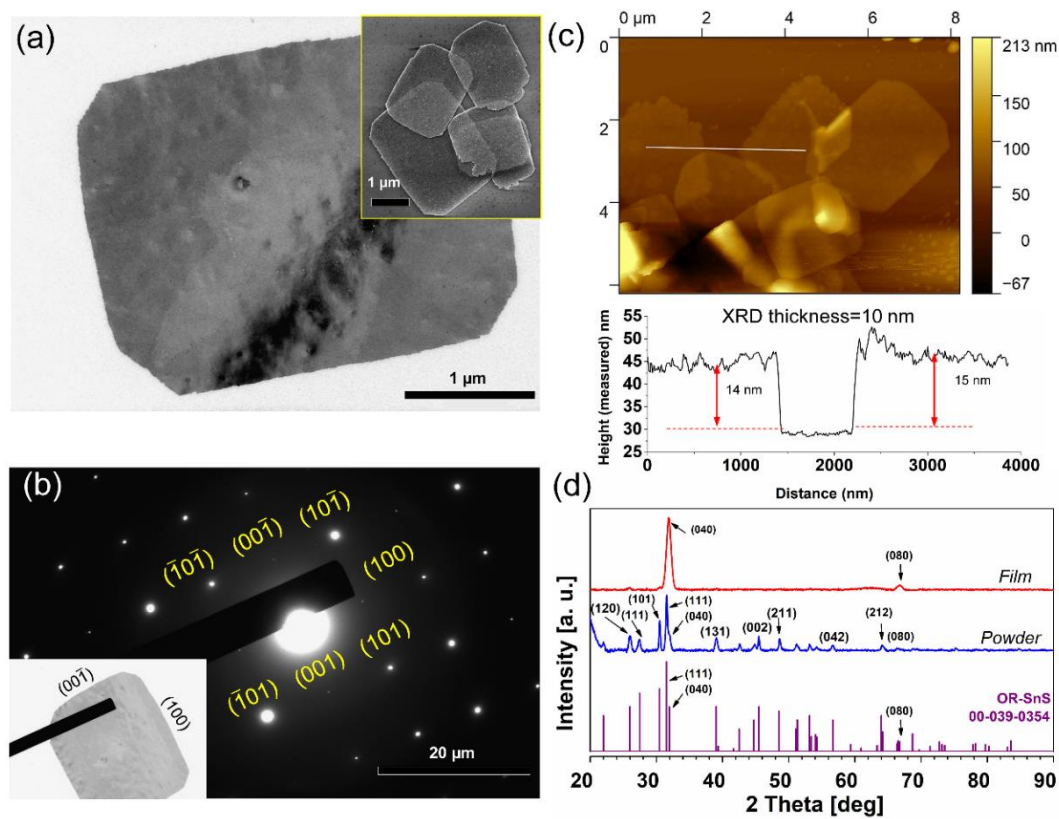

**Figure 1.** (a) TEM image and SAED pattern (b) of SnS NSs. (c) AFM image and height profile of two single SnS nanosheets. (d) XRD patterns for SnS NSs drop-casted on a Si wafer (thin film) and a powder sample in a capillary tube.

Recently, we investigated SnS NSs synthesized using thioacetamide (TAA) as sulfur source and tuning the oleic acid/trioctylphosphine (OA/TOP) ligand ratio to get NSs in different shapes (between squared and hexagonal shapes) and various thicknesses (~20-50 nm).[16] In this new approach, we adopt TOP-S as sulfur source as an improved recipe. A typical transmission electron microscope (TEM) image of a standard sample of SnS NSs with lateral sizes of approximately 3100 x 2500 nm is shown in Figure 1a. A representative scanning electron microscope (SEM) image of dispersed NSs is shown as inset (Figure 1a), presenting well defined rectangular nanosheet morphology. The selected area electron diffraction pattern (SAED) demonstrates a spot pattern that is consistent with a single-crystal nanosheet (Figure 1b). The diffraction spots of the SAED can be indexed to the specific facets of OR-SnS, which are also identified in the corresponding TEM image in the inset of Figure 1b (SAED patterns are rotated with respect to the TEM image by instrumentation). Atomic force microscopy (AFM, Figure 1c) reveals that the thickness of a single sheet is around 14-15 nm, and the calculated average thickness with XRD data using the Scherrer equation is 10 nm. Powder X-ray diffraction (XRD, Figure 1d) for the capillary sample matches well the peaks from the reference card (00-039-0354) for OR SnS. The XRD pattern of a drop-casted sample shows prominent preferred orientation due to the texture effect (the large lateral size compared to the small vertical thickness makes crystallographic orientations not random), with significantly pronounced (040), (080) reflexes. This indicates that the thickness direction is [010], showing highly textured orientation for drop-casted samples. This is consistent with the SAED suggesting that the surface is oriented along [010] direction (Figure 1b). Both XRD patterns confirm the pure OR crystal structure. The optical band gap ($E_g$) of the SnS NSs (a-SnS) is determined by the Tauc linearization based on the optical absorption measurements (Figure S1). The values for the direct and indirect bandgaps of OR-SnS NSs are determined to be 1.69 eV and 1.25 eV, respectively. Varying the reaction parameters to 0.52 mmol 2M TOP-S for the injection, we successfully obtained the thinnest SnS NSs in 2.4 μm large. (7.0 nm from XRD data, 11 nm from AFM data, Figure S2) after 30s reaction.

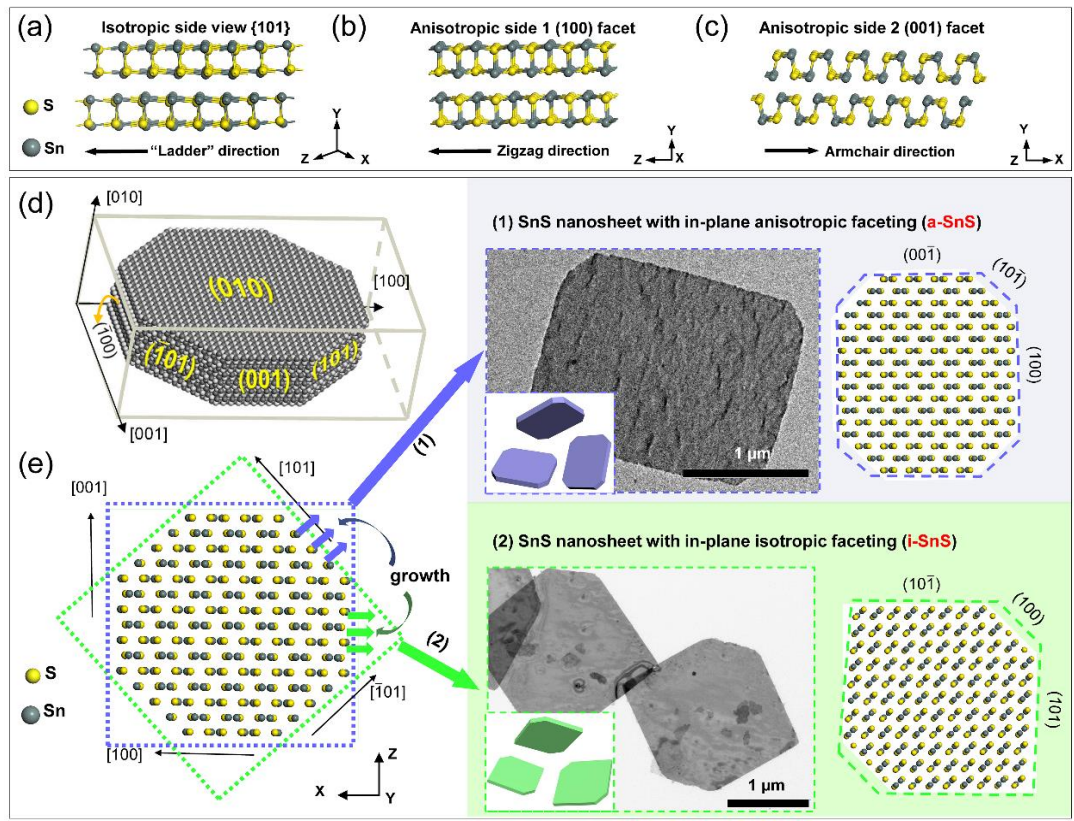

**Figure 2.** The atomic arrangements of exposed facets parallel to the "ladder" direction (a), zigzag direction (b) and armchair direction (c) for SnS NSs. (d) A 3D crystal model of an octagonal prism of SnS with observable planes marked. (e) Morphology and corresponding atomic models of SnS NSs with exposed facets parallel to the anisotropic directions (forming a-SnS) or isotropic directions (forming i-SnS). These two types of NSs grow respectively from the same original starting point with different growth directions (shown in blue and green arrows).

Recently, in-plane electronic anisotropy of 2D materials has gained great attentions and it has been widely investigated for potential electronic devices.[5, 17-18] Therefore, in-solution tuning of edge faceting plays an important role in the application of electronic anisotropy for 2D colloidal nanomaterials. Generally, after adjusting reaction parameters, we achieved two different types of SnS NSs in respect to edge faceting: 1. Rectangular NSs with exposed edges parallel to the (100), (001) planes (parallel to zigzag and armchair direction, respectively); 2. Rectangular (nearly square-shaped) NSs with edges parallel respectively to the {101} planes, which are along four isotropic "ladder" directions (Figure 2). We define them as anisotropic-SnS (a-SnS) and isotropic-SnS (i-SnS) NSs respectively. The crystal structure of SnS is shown in Figure

2. It is an orthorhombic crystal structure with Sn and S atoms in two adjacent double layers, and layers stacking along the Y (or b) axis by weak van der Waals interaction.[6] A 3D crystal model of an octagonal prism is shown (Figure 2d) as simplified starting point for investigating further growth of shaped nanocrystals. The a-SnS NSs are obtained by using 0.64 mmol OA and 1.0 mmol TOP in a flask, followed by the hot injection of 0.26 mL 1M TOP-S at 300 ºC. The obtained SnS NSs are in rectangular shape with edges parallel to the armchair/zigzag directions, according to the interpretation of the SAED in Figure S3 and Figure 2e-(1).

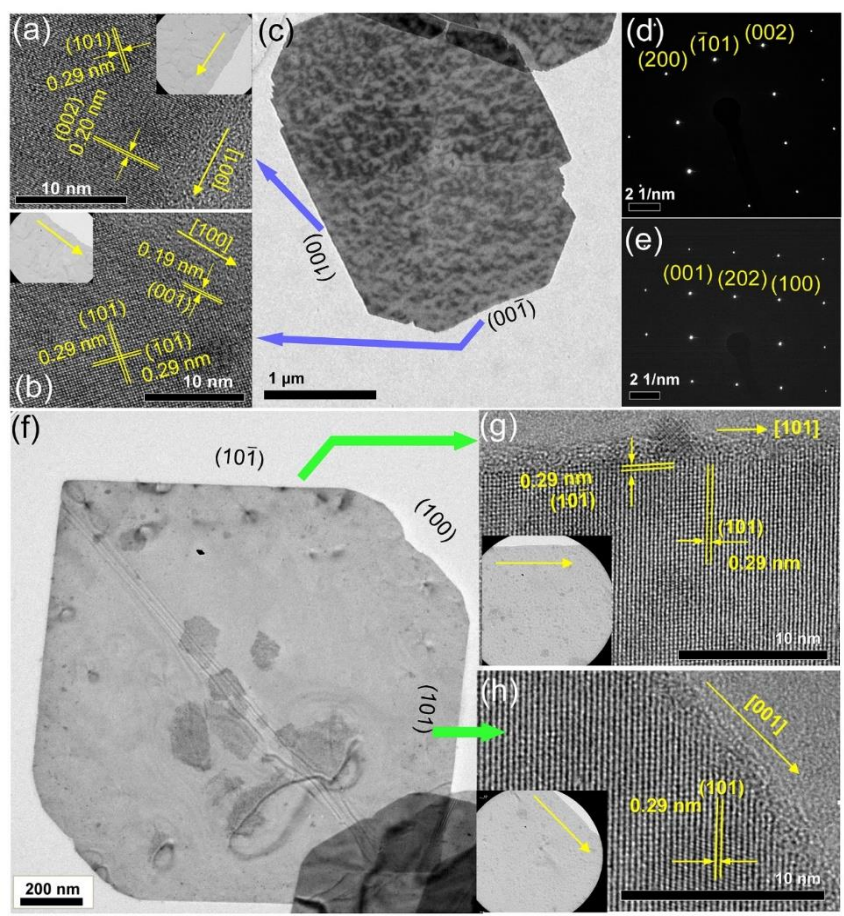

**Figure 3.** High-resolution TEM images and SAED patterns of rectangular a-SnS NSs (a, b, c, d) and i-SnS NSs (e, f, g, h).

The HRTEM was performed to confirm the edge faceting (Figure 3). The lattice fringes with the *d*-spacing of 0.20 nm for the (002) plane and the SAED pattern (Figure 3a, d) indicates the longer rectangular edge of a-SnS has exposed (100) facets, whereas the lattice spacing of 0.19 nm for the (00$\bar{2}$) plane and in-plane nearly perpendicular planes with both *d*-spacing of 0.29 nm (($10\bar{1}$) and ($\bar{1}0\bar{1}$) planes) in Figure 3b demonstrates the shorter rectangular edge is parallel to ($00\bar{1}$) planes, proven also by the SAED (Figure 3d). Additionally, it also has four truncated corners parallel to the {101} planes. The other type, i-SnS NSs, is prepared with 1.56 mmol OA and hot injection of 0.52 mL 1M TOP-S. The synthesized sheets are rectangular SnS NSs with four main edges parallel to the "ladder" directions and a pair of truncated corners parallel to (100) facets according to SAED in Figure S3b and Figure 2e-(2). Likewise, the SAED and the *d*-spacing of 0.29 nm (Figure 3e, g, h) confirmed that the four main edges are parallel to the exposed {101} planes and the two truncated corner edges correspond to the exposed (100) facets (Figure S3b).

In the previous work, we have already investigated the adsorption energies of capping ligands and their role in anisotropic growth of SnS square and hexagonal shaped NSs.[16] Here, we demonstrate the anisotropic growth of nanocrystals under different reaction conditions to form two types of sheets (or anisotropic faceting). After taking aliquots to check the growth process (Figure S4), we confirmed a similar growth mechanism as the NSs reported recently.[16] Thus, the anisotropic growth governed by capping ligands after fast nucleation still serves as the main growth pattern for the SnS NSs investigated in this study. The variation of capping ligands in the reaction plays an important role in tuning the growth anisotropy of SnS NSs. Thus, we investigated the influence of the amount of TOP (in the flask and in the TOP-S precursor solution) and OA respectively (Figure S5-S7). More TOP-S used in the reaction can still facilitate the formation of larger and thicker a-SnS NSs (Figure S5a-c, 0.26 mL, 0.52 mL, 0.78 mL 1M TOP-S). The thicknesses are 10, 12, 36 nm determined from XRD data and each corresponding AFM thickness is about 5~7 nm larger compared to the XRD thickness (partly shown in Figure S6). Interestingly, if TOP only serves as solvent for dissolving S (TOP-S), with 0.64 mmol OA, only undefined-shaped NSs are formed (Figure S5d). Increasing amount of TOP-S, a-SnS NSs start to appear composed of four explicit rectangular edges (even no truncated corners formed, Figure S5e). The lateral size decreases from nearly 8 to 3.5 μm, whereas the thickness fluctuates from 8.9 to 9.3 nm according to the XRD data (Figure S5f). A similar tendency can also be observed when we raised the amount of OA without any TOP in the flask before injection (Figure S7). The square NSs were produced with a higher amount of OA (from 0.64 to 1.5 and 2.0 mmol) in the reaction. Nevertheless, edge faceting switches to exposed facets parallel to isotropic directions, forming i-SnS NSs with 1.5 and 2.0 mmol OA (Figure S7b, c). In terms of morphology, the lateral size becomes smaller upon the increase of the OA amount (respectively 7900 to 1500 and 1400 nm) and the thickness becomes larger (respectively 8.9 to 17.6 and 27 nm, defined by XRD, Figure S7d, corresponding AFM images partly shown in Figure S8, S9). When 1.0 mmol TOP is introduced with a high amount of OA (1.5 mmol), i-SnS NSs are also formed, whereby truncated edges which correspond to (100) and ($\bar{1}$00) crystal planes, are more obvious (Figure S10). Theoretically, binding affinities of specific ligands on preferred facets play a significant role in anisotropic growth for nanocrystals.[19-20] The a-SnS NSs are formed by a higher growth speed on {101} facets, whereby the growth on (100), (001) and their parallel facets ($\bar{1}$00) and (00$\bar{1}$), is hampered due to stronger passivation of OA.[16] Then, i-SnS NSs are formed with relatively faster growth towards the [100] (or [$\bar{1}$00]) direction, and [001] (or [00$\bar{1}$]) directions, coupled with a passivation of the {101} facets. To support this observation, DFT simulations are performed to investigate the possible mechanism for the formation of the edge faceting based on the adsorption energy of ligands on each representative facet of SnS nanocrystals (Table 1). Compared to our previous work using TAA for the i-SnS NS synthesis, TOP-S is considered as the key factor for obtaining a-SnS NSs. In the case of TAA for the SnS NS synthesis, i-SnS NSs can be prepared with the co-passivation of applied ligands (oleate, TOP and OA), due to the high adsorption energy of TOP and oleate on {101} and (100) facets.[16] Here, the formation of a-SnS is mainly due to the high adsorption energy of TOP-S on (100), (001) facets (DFT, Table 1). Higher adsorption energy of TOP-S may sterically hinder the approaching of SnS monomers to the surface, subsequently slowing the growth speed of the facets. In detail, the presence of TOP-S in the reaction solution obviously exert a strong influence on the SnS crystal surface, preferentially stabilizing the (100), (001) facets. This finally impacts on the growth rate of these facets during growth, so it's rational to propose the (100) and (001) facets turn to be more stable and present lower growth rate, which in the end serve as the exposed facets for the final sheets. It finally results in specific faceting with edges along anisotropic directions.[21] The simulations of TOP-S on the facets using simplified molecules with different chain lengths also show similar tendencies (Table S1). What is more fascinating is that the in-plane faceting can be switched from a-SnS to i-SnS by adjusting the amount of OA in the reaction even in the presence of TOP-S. This is also supported by the DFT results in our previous work[16], exhibiting high absorption energy of oleate on (100) and {101} facets, leading to square-shaped sheets with (100) truncated corner edges in the end.

**Table 1.** Adsorption energy [eV] of ligand molecules on the four facets of SnS calculated by the density functional theory (DFT) method. The simulations were performed using simplified molecules for TOP, TOP-S (tripropylphosphine (TPP), and TPP-S).

|            | SnS-101 side facet (isotropic) | SnS-100 side facet (anisotropic-zigzag) | SnS-001 side facet (anisotropic-armchair) | SnS-010 top facet (Top or down) |
|------------|-------------------------------|-----------------------------------------|-------------------------------------------|--------------------------------|
| **TPP**    | 1.939                         | 1.503                                   | 1.551                                     | 0.675                          |
| **TPP-S (C3)** | 1.740                     | 2.390                                   | 1.799                                     | 0.927                          |

The zigzag and armchair directions have been investigated to demonstrate anisotropy in electronic behavior based on the difference in atomic ordering along these two directions.[6] To investigate the electrical properties of these NSs along different crystallographic directions, we contacted them individually and measured their room temperature conductivity. In one group of the devices, four contacts were placed parallel to the edges of the NSs in order to investigate the conductivity along the anisotropic directions (Figure 4a). For the other group, the contacts were placed from corner to corner to probe the isotropic directions (Figure 4b). Figure 4c shows the IV characteristics of these sheets along the anisotropic directions. The conductivity along the (100) direction (zigzag) reaches 65 S/m while for the (001) direction (armchair), a significantly lower conductivity of 39 S/m can be observed. The ratio of the conductivity along these directions (zigzag/armchair) is equal to 1.67, which is in very good agreement with the literature.[6] The effective mass of the carriers, which are mainly holes[6, 22], differs considerably along these two directions.[6, 22-23] As a result of this difference, the carriers have a higher mobility along the zigzag direction, which results in a higher conductivity.[6] On the other hand, the conductivity was measured along the (101), (10$\bar{1}$) directions (ladder directions). The conductivity along these directions is isotropic (2.2 S/m and 2.3 S/m, Figure 4d). The arrangement of atoms along these directions is exactly identical which results in a comparable electrical conductivity.

In summary, we investigated the crystal-structure orientation switching in respect to the edge morphology of rectangular-shaped SnS NSs. We report the synthesis conditions when the edge faceting can be switched between faceting along two anisotropic directions and along four isotropic directions. The tuning of the thickness (down to 7 nm) and lateral size (up to 8 μm) for SnS NSs manifests a promising advantage of these materials. The synthesis opens the possibility to contact individual nanosheet on the substrate, with predefined orientations, promoting the applications of colloidal 2D nanomaterials for new devices with anisotropic properties. We exemplify this by electrical measurements. They confirm different conductivities for crystallographically distinct direction. It displays an in-plane electronic anisotropy in conductivity along the zigzag and armchair directions, but isotropic behavior from corner to corner (ladder directions).

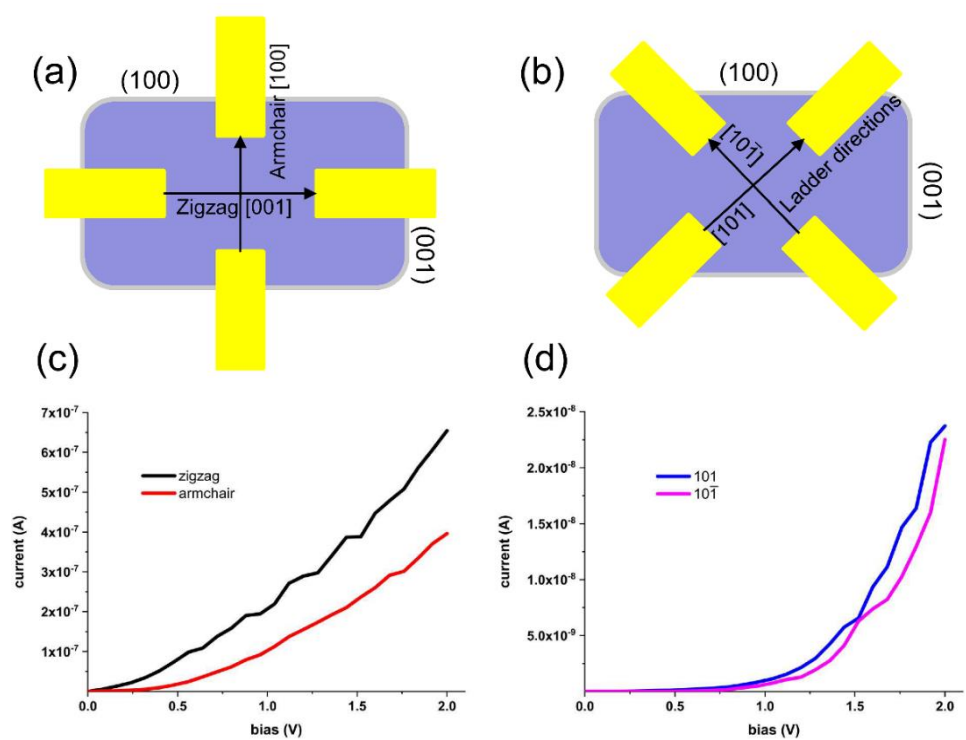

**Figure 4.** Electrical measurements along different directions. Schematic illustration of the nanosheets, contacted along anisotropic directions (a) and along isotropic directions (b). (c) The IV characteristics for anisotropic directions showing a higher conductivity along the zigzag direction compared to the armchair direction (d) The IV characteristics along isotropic directions, representing similar conductivities along the ladder directions.

**Experimental Section**

*Preparation and Characterization of 2D SnS Nanosheets:* The synthesis and characterization methods are detailly described in the supporting information.

*Device preparation and transport measurements:* the individual SnS nanosheet was contacted by e-beam lithography and measured by a semiconductor parameter analyzer (detail in the supporting information).

**Associated Content**

Supporting Information is available free of charge on the ACS Publications website at DOI:

## Acknowledgements

This work was supported by China Scholarship Council (CSC). We thank the European Research Council for the ERC Starting Grant "2D-SYNETRA" (Seventh Framework Program FP7, Project: 304980) and the German Research Foundation DFG for financial support in the frame of the Cluster of Excellence "Center of ultrafast imaging CUI" and the Heisenberg scholarship KL 1453/9-2. We thank Daniela Weinert and Almut Barck (Universität Hamburg) for help with the HRTEM and XRD measurements.